\documentclass[preprintnumbers,article,amsmath,amssymb,floatfix,10pt,prd,onecolumn,superscriptaddress,nofootinbib]{revtex4-2}

\usepackage{titlesec}
\usepackage[toc]{appendix}

\titleformat*{\section}{\LARGE\bfseries}
\titleformat*{\subsection}{\Large\bfseries}
\titleformat*{\subsubsection}{\large\bfseries}
\titleformat*{\paragraph}{\large\bfseries}
\titleformat*{\subparagraph}{\large\bfseries}
\usepackage{bm}
\usepackage{amsfonts}
\usepackage{latexsym}
\usepackage[latin1]{inputenc}
\usepackage{graphicx}
\usepackage{amsmath}
\usepackage{palatino}
\usepackage{mathpazo}
\usepackage{textcomp}
\usepackage{float}
\usepackage{booktabs}
\usepackage{dcolumn}
\usepackage{ragged2e}
\usepackage{hyperref}
\hypersetup{colorlinks,citecolor=blue}
\hypersetup{colorlinks=true,linkcolor=red,filecolor=magenta,    urlcolor=blue}
\usepackage{amsthm}
\usepackage{xcolor}
\usepackage{orcidlink}
\usepackage{epsfig}
\usepackage[caption=false]{subfig}
\usepackage{commath}
\usepackage{cancel, ulem}

\usepackage{multirow}
\newtheorem{remark}{Remark}
\newcommand{\m}{\mathring}
\newcommand{\be}{\begin{equation}}
\newcommand{\ee}{\end{equation}}
\newcommand{\bea}{\begin{eqnarray}}
\newcommand{\eea}{\end{eqnarray}}
\newcommand{\eeas}{\end{eqnarray*}}
\newcommand{\beas}{\begin{eqnarray*}}

\def\jnl@style{\it}
\def\aaref@jnl#1{{\jnl@style#1}}

\def\aaref@jnl#1{{\jnl@style#1}}

\def\aj{\aaref@jnl{AJ}}                   
\def\apj{\aaref@jnl{ApJ}}                 
\def\apjl{\aaref@jnl{ApJ}}                
\def\apjs{\aaref@jnl{ApJS}}               
\def\apss{\aaref@jnl{Ap\&SS}}             
\def\aap{\aaref@jnl{A\&A}}                
\def\aapr{\aaref@jnl{A\&A~Rev.}}          
\def\aaps{\aaref@jnl{A\&AS}}              
\def\mnras{\aaref@jnl{Mon.~Not.~Roy.~Astron.~Soc.}}             
\def\prd{\aaref@jnl{Phys.~Rev.~D}}        
\def\prc{\aaref@jnl{Phys.~Rev.~C}}  
\def\prl{\aaref@jnl{Phys.~Rev.~Lett.}}    
\def\qjras{\aaref@jnl{QJRAS}}             
\def\skytel{\aaref@jnl{S\&T}}             
\def\ssr{\aaref@jnl{Space~Sci.~Rev.}}     
\def\zap{\aaref@jnl{ZAp}}                 
\def\nat{\aaref@jnl{Nature}}              
\def\aplett{\aaref@jnl{Astrophys.~Lett.}} 
\def\apspr{\aaref@jnl{Astrophys.~Space~Phys.~Res.}} 
\def\physrep{\aaref@jnl{Phys.~Rep.}}      
\def\physscr{\aaref@jnl{Phys.~Scr}}       
\def\commat{\aaref@jnl{Comm.~Math.~Phys.}}              
\def\science{\aaref@jnl{Science}}               
\def\cqg{\aaref@jnl{Classical Quant.~Grav.}}            
\def\jpcs{\aaref@jnl{JPCS}}                                     
\def\ijmpd{\aaref@jnl{Int.~J.~Mod.~Phys.~D}}                    
\def\grg{\aaref@jnl{Gen.~Relat.~Gravit.}}               
\def\rpp{\aaref@jnl{Rep.~Prog.~Phys.}}          
\def\npa{\aaref@jnl{Nucl.~Phys.~A}}        
\def\lrr{\aaref@jnl{Living Rev.~Rel.}}                   
\def\jcap{\aaref@jnl{J.~Cosmology Astropart.~Phys.}}    
\def\rmp{\aaref@jnl{Rev.~Mod.~Phys.}}   
\def\epjc{\aaref@jnl{Eur.~Phys.~J.~C}} 
\def\plb{\aaref@jnl{~Phy.~Lett.~B}} 
\def\mpla{\aaref@jnl{Mod.~Phy.~Lett.~A}} 
\def\arxiv{\aaref@jnl{arxiv.org}}

\allowdisplaybreaks[1]

\begin{document}

\title{On the viability of $f(Q)$ gravity models}
\author{Avik De\orcidlink{0000-0001-6475-3085}}
\email{de.math@gmail.com}
\affiliation{Department of Mathematical and Actuarial Sciences, Universiti Tunku Abdul Rahman, Jalan Sungai Long,
43000 Cheras, Malaysia}
\author{Tee-How Loo\orcidlink{0000-0003-4099-9843}}
\email{looth@um.edu.my}
\affiliation{Institute of Mathematical Sciences, Faculty of Science, Universiti Malaya, 50603 Kuala Lumpur, Malaysia}


\footnotetext{The research was supported by the Ministry of Higher Education (MoHE), through the Fundamental Research Grant Scheme (FRGS/1/2021/STG06/UTAR/02/1). }

\begin{abstract}
In general relativity, the contracted Bianchi identity makes the field equation compatible with the energy conservation, likewise in $f(R)$ theories of gravity. We show that this classical phenomenon is not guaranteed in the symmetric teleparallel theory, and rather generally $f(Q)$ model specific. We further prove that the energy conservation criterion is equivalent to the affine connection's field equation of $f(Q)$ theory, and except the $f(Q)=\alpha Q+\beta$ model, the non-linear $f(Q)$ models do not satisfy the energy conservation or, equivalently the second field equation in every spacetime geometry; unless $Q$ itself is a constant. So the problem is deep-rooted in the theory, several physically motivated examples are provided in the support.     
\end{abstract}

\maketitle

\section{\textbf{Introduction}}

Although Einstein's general relativity (GR) has undeniably been successful, its shortcomings in a number of areas have become apparent in recent years, prompting academics to explore alternatives. 
Later on, in general a metric-compatible affine connection on flat spacetime with torsion was substituted for the unique torsion-free and metric-compatible Levi-Civita connection 
on which GR was originally built, and its torsion was allowed to shoulder the whole burden of characterising gravity. This particular theory, initiated by Einstein himself \cite{einst}, is called the metric teleparallel gravity. Recently, a newborn has emerged in this clan, the symmetric teleparallel theory, formulated based on an affine connection with vanishing curvature and torsion, and gravity was attributed to the non-metricity of the spacetime \cite{nester}. One can construct the so-called torsion scalar $\mathbb{T}$ from the torsion tensor in the metric teleparallel theory and the non-metricity scalar $Q$ in its symmetric counterpart. Thereafter, by considering the Lagrangian $\mathcal{L}=\sqrt{-g}\mathbb{T}$ in the former and $\mathcal{L}=\sqrt{-g}Q$ in the latter, the respective field equations can be obtained. 
However, the both of these theories are equivalent to GR up to a boundary term since both the scalars $\mathbb{T}$ and $Q$ are equal to the Levi-Civita Ricci scalar modulo a  surface term. Naturally, both the metric and symmetric teleparallelism inherit the same `dark' problem as in GR. To address this issue, modified $f(\mathbb{T})$ \cite{f(T)} and $f(Q)$ \cite{coincident} theories of gravity in the respective genre have been introduced. It is important to keep in mind that, unlike GR, in the teleparallel theory the affine connection is independent of the metric tensor, and so  the teleparallel theories are technically metric-affine theories where both the metric and the connection act as dynamic variables. 

Both $f(\mathbb{T})$ and $f(Q)$ theories appear in second order field equation, as opposed to fourth order ones in $f(R)$, providing an immediate advantage over $f(R)$. Both of these theories have the potential to offer an alternative rationale for the accelerating expansion of the Universe 
\cite{accfT1, accfT2, accfT3, accfT4, dy2, accfQ1, accfQ2, accfQ3}. However, models based on $f(\mathbb{T})$ suffer from considerable coupling issues and local Lorentz invariance problem, which is not a concern for $f(Q)$ theories. For a deeper comparison of these two interesting branches of the teleparallel theories of gravity, see \cite{fQfT,fQfT1,fQfT2,fQfT3} and the references therein.

In GR the contracted Bianchi identities validate the energy conservation law of ordinary matter, and it profoundly separates the geometrical and matter sectors in this standard theory of gravity, otherwise known as minimal matter-curvature coupling. The theory of GR is formulated based on few fundamental principles, the law of energy conservation is one of them. The null divergence of the left hand side, i.e., the gravitational part of Einstein's original field equation in GR is straightforwardly obtained by contracted Bianchi identities, and thus the field equation are compatible with the classical law of energy conservation, that is, null divergence of the energy-momentum tensor. So a natural question arises when some new gravity theory is introduced, whether the energy-momentum is covariantly conserved or not \cite{conserve,conserve1,conserve2,conserve3,cosmoconserve}; the symmetric teleparallel theories are no exception. In the present study, we illustrate that the energy conservation compatibility is not a trivial issue to deal with in $f(Q)$ theories, unlike in GR and also in $f(R)$ where straightforward calculation can yield this, independent of models chosen. On the contrary, this law can be employed to filter the possible $f(Q)$ gravity models. We term the vanishing of covariant divergence of the left hand side of any modified gravity theories, the extended Bianchi identity. This issue is also present in the $f(\mathbb{T})$ theories \cite{bianchift,bianchift2,fT/issue,
fT/issue2,fT/lli,pereira,krssak,krssak2,bad-tetrad,weinberg,blagojevi}, and we observe that the anti-symmetric part of the field equation of $f(\mathbb{T})$ obtained from the variation of the action term with respect to the metric tensor actually equals the field equation obtained from the variation of the same action by the affine connection. So only when the connection's field equation in $f(\mathbb{T})$ is fulfilled by some model in some underlying geometry, the anti-symmetric part of the metric field equation vanishes and it in turn confirms that the energy conservation is satisfied by that $f(\mathbb{T})$ model. Consequently, the field equation of $f(\mathbb{T})$ theory is not always compatible with the classical energy conservation. Unfortunately, this crucial feature of the $f(\mathbb{T})$ theory is mostly ignored in the literature since the connection field equation vanishes predominantly in the isotropic and homogeneous Friedmann-Lema\^itre-Robertson-Walker (FLRW) spacetime. We demonstrate in our present study that this luxury of ignorance is absent in the arena of symmetric teleparallel theory of gravity, even in the spatially flat FLRW spacetime. 

The article is organized as follows:\\
After the introduction, we set the stage in Section \ref{sec2} for teleparallel geometry; followed by a detailed analysis of $f(Q)$ theories of gravity in Section \ref{sec3}. The section appears with $f(Q)$ action principle, two field equations corresponding to the variation with respect to metric and connection, the extended Bianchi identity, and the equivalence between the energy conservation and the connection's field equation. We continue to offer a condition to check the compatibility of energy condition criterion in $f(Q)$ theory and provide investigation of some important ansatz in cosmological and astrophysical research. We finally present our conclusion in Section \ref{sec5}.

Throughout the article we have used the notations $f_Q=\frac{df}{dQ}, \quad f_{QQ}=\frac{d^2f}{dQ^2}$. All the expressions with a $\mathring{(~)}$ is calculated with respect to the Levi-Civita connection $\mathring{\Gamma}$.

\section{\textbf{The formulation of teleparallel theory}}\label{sec2}

A brief review on the geometric formulation of the teleparallel theory is given in this section. Consider a spacetime equipped with a metric tensor $g_{\mu\nu}$ and an affine connection $\Gamma^\lambda{}_{\mu\nu}$. 
The torsion tensor $T^\lambda_{}{\mu\nu}$ and non-metricity tensor $Q_{\lambda\mu\nu}$ are given respectively as 
\begin{align}
T^\lambda_{}{\mu\nu}&=-\Gamma^\lambda{}_{\mu\nu}+\Gamma^\lambda{}_{\nu\mu}\,,\\
Q_{\lambda\mu\nu}&=\partial_\lambda g_{\mu\nu}-\Gamma^{\sigma}{}_{\mu\lambda}g_{\sigma\nu}
-\Gamma^{\sigma}{}_{\nu\lambda}g_{\sigma\mu}\equiv\nabla_\lambda g_{\mu\nu}.
\end{align}
The relation between the affine connection and the Levi-Civita connection $\mathring{\Gamma}^\lambda{}_{\mu\nu}$ is given by
\begin{equation} \label{connc}
\Gamma^\lambda{}_{\mu\nu} = \mathring{\Gamma}^\lambda{}_{\mu\nu}+M^\lambda{}_{\mu\nu}\,.
\end{equation}
The tensor $M^\lambda{}_{\mu\nu}$ is known as the distortion tensor, which can be decomposed as 
\begin{align}
M^\lambda{}_{\mu\nu}=L^\lambda{}_{\mu\nu}+K^\lambda{}_{\mu\nu}\,,
\end{align}
where 
\begin{align}\label{L}
L^\lambda{}_{\mu\nu} = \frac{1}{2} (Q^\lambda{}_{\mu\nu} - Q_\mu{}^\lambda{}_\nu - Q_\nu{}^\lambda{}_\mu) \,,
\end{align} 
is the disformation tensor and 
\begin{align}\label{K}
K^\lambda{}_{\mu\nu} = -\frac{1}{2} (T^\lambda{}_{\mu\nu} - T_\mu{}^\lambda{}_\nu - T_\nu{}^\lambda{}_\mu) \,,
\end{align} 
is the contorsion tensor.
The curvature tensor corresponding to $\Gamma^\lambda{}_{\mu\nu}$ is related to that of 
$\mathring{\Gamma}^\lambda{}_{\mu\nu}$ by
\begin{align}\label{curvature}
R^\lambda{}_{\mu\alpha\nu}=\mathring R^\lambda{}_{\mu\alpha\nu}
+\mathring\nabla_\alpha M^\lambda{}_{\mu\nu}
-\mathring\nabla_\nu M^\lambda{}_{\mu\alpha}
+M^\lambda{}_{\sigma\alpha}M^\sigma{}_{\mu\nu}
-M^\lambda{}_{\sigma\nu}M^\sigma{}_{\mu\alpha}\,.
\end{align}

Two of the special cases, namely the metric teleparallel and symmetric teleparallel theories are of particular interest to generalize the Levi-Civita connection based theory of GR as well as its extension, $f(R)$ theory. In contrast, the curvature free and vanishing non-metricity conditions are imposed for the metric teleparallel theory while the affine connection is constrained to be torsionless and of curvature free in the case of symmetric teleparallelism.


\section{\textbf{Energy conservation in $f(Q)$ theories of gravity}}\label{sec3}
By all means, theories of gravity based on curvature and torsion are almost in their adult age, whereas theories on the non-metricity is in a very initial stage and requires a lot of scrutiny and detailed analysis, in their formulation, theoretical viability and observational support. Several important publications came up very recently on this modified $f(Q)$ gravity theory and its cosmological implications, see \cite{lcdm,deepjc,zhao,gde,lin,cosmography,signa,redshift,perturb,
dynamical1,dynamical2,latetime,quantum, bouncing,bigbang,Avik/prd,FLRW/connection1} and the references therein. However, the important issue of energy conservation in the non-metricity based symmetric teleparallel gravity was surprisingly neglected in the previous literature, as was its second field equation obtained from the variation of the action term with respect to the affine connection.

In the modified $f(Q)$ theories of gravity under symmetric teleparallelism, we consider an affine connection $\Gamma^\lambda{}_{\mu\nu}$ 
with vanishing curvature and null torsion and let the non-metricity alone control the gravity. Since the torsion $T^\lambda{}_{\mu\nu}=0$, the tensor $M^\lambda{}_{\mu\nu}$ reduces to $L^\lambda{}_{\mu\nu}$ in the present scenario.  
We define the non-metricity scalar  \cite{coincident} 
\begin{align} \label{Q_0}
Q=Q_{\lambda\mu\nu}P^{\lambda\mu\nu}= \frac{1}{4}(-Q_{\lambda\mu\nu}Q^{\lambda\mu\nu} + 2Q_{\lambda\mu\nu}Q^{\mu\lambda\nu} +Q_\lambda Q^\lambda -2Q_\lambda \tilde{Q}^\lambda)\,,
\end{align}
where the tensor $P^\lambda{}_{\mu\nu}$ is given by
\begin{equation} \label{P}
P^\lambda{}_{\mu\nu} = \frac{1}{4} \left( -2 L^\lambda{}_{\mu\nu} + Q^\lambda g_{\mu\nu} - \tilde{Q}^\lambda g_{\mu\nu} -\frac{1}{2} \delta^\lambda_\mu Q_{\nu} - \frac{1}{2} \delta^\lambda_\nu Q_{\mu} \right) \,,
\end{equation} 
$Q_{\lambda}=Q_{\lambda\mu\nu}g^{\mu\nu}, \, \tilde Q_{\nu}=Q_{\lambda\mu\nu}g^{\lambda\mu}$ are the two possible types of traces of the non-metricity tensor. 
The non-metricity scalar $Q$ is invariant under local general linear transformations and as well as translational symmetries.
When the coincident gauge is taken in which the affine connection has vanishing components, the scalar $Q$ can be expressed as
\begin{align}
Q=g^{\mu\nu}(\m\Gamma^\alpha{}_{\sigma\nu}\m\Gamma^\sigma{}_{\mu\alpha}
-\m\Gamma^\alpha{}_{\sigma\alpha}\m\Gamma^\sigma{}_{\mu\nu})\,,
\end{align}
which appeared to be the kernel of a Lagrangian  density associated to the prototype of the $Q$-theory theory that was frequently adopted in the past; for instance, the derivation of energy-momentum density for gravity in \cite{tolman}.

The action of $f(Q)$ theory is decomposed into its gravitational and matter parts as  
$S=S_G+S_M$, where $S_G=\frac1{2\kappa}\int f(Q) \sqrt{-g}\,d^4 x$ is the gravitational action and $S_M=\int \mathcal{L}_M \sqrt{-g}\,d^4 x$ is the matter action. The gravitational action depends on the metric and the affine connection generically, however, we assume the matter action $S_M$ to be dependent only on the metric and some matter fields $\xi^A$, the affine connection is not involved.

As the affine connection is also a dynamic variable for the action term, in contrast to GR, there is another supplemented  field equation attributed to the affine connection in addition to the metric field equation.
Noticing that the variation of the non-metricity reads as 
\begin{align}\label{var_Q}
\delta Q_{\lambda\mu\nu}=\nabla_\lambda\delta g_{\mu\nu}
-g_{\sigma\nu}\delta\Gamma^\sigma{}_{\mu\lambda}-g_{\mu\sigma}\delta\Gamma^\sigma{}_{\nu\lambda}\,.
\end{align}
It follows from (\ref{var_Q}) that the variation of the  gravitational action takes the form
\begin{align}\label{var_S_G}
\delta S_G=-\frac1{2\kappa}\int (E^{\mu\nu}\delta g_{\mu\nu}
+4f_Q P^{\nu\mu}{}_\lambda\delta\Gamma^\lambda{}_{\mu\nu}
)\sqrt{-g}\,d^4 x\,,
\end{align}
where 
\begin{align}\label{E-1}
E^{\mu\nu}
=&\frac{2}{\sqrt{-g}} \partial_\lambda (\sqrt{-g}f_QP^{\lambda\mu\nu}) -\frac{1}{2}f g^{\mu\nu} 
        + f_Q(P^{\mu\sigma\rho}Q^{\nu}{}_{\rho\sigma}+2P^{\sigma\rho\mu}Q_{\sigma\rho}{}^\nu)\,.
\end{align}
Hence one can easily derive the variation of the total action with respect to the metric, which reads
\[
\delta_g S=\frac1{2\kappa}\int (-E^{\mu\nu}+\kappa\Theta^{\mu\nu})\delta g_{\mu\nu}\sqrt{-g}\,d^4 x\,,
\]
where $\Theta^{\mu\nu}=\frac2{\sqrt{-g}}\frac{\delta(\sqrt{-g}L_M)}{\delta g_{\mu\nu}}$ is the energy-momentum tensor. The corresponding metric field equation 
\begin{align}\label{FE1}
E^{\mu\nu}=\kappa\Theta^{\mu\nu}\,,
\end{align}
is then deduced due to the covariant invariance of the total action. Varying the action term with respect to the affine connection, we obtain the other field equation of $f(Q)$ theory 
\begin{align}\label{FE2}
\partial_\mu\partial_\nu(\sqrt{-g}f_Q P^{\nu\mu}{}_\lambda)=0\,.
\end{align}
In this context, it is utterly necessary to bring up the crucial point that not all $f(Q)$ models satisfy (\ref{FE2}) for a given spacetime geometry, to be shown in a later stage. What follows next is a result showing explicitly that the affine connection's field equation (\ref{FE2}) is equivalent to the classical energy-conservation. In particular,
\begin{align}\label{equiv}
\frac2{\sqrt{-g}}\partial_\lambda\partial_\alpha\left(\sqrt{-g}f_QP^{\alpha\lambda}{}_{\nu}\right)
=\kappa\m\nabla_\mu \Theta^\mu{}_\nu\,.
\end{align}
For the complexity of the proof, we provide the detailed steps of it in Appendix \ref{appEquiv}. This shows that the energy conservation, that is, the divergence-free property of the energy-momentum tensor, is not only a physically reasonable additional requirement, but actually a consequence of the theory itself.

Now, let us concentrate on the covariant divergence of the energy-momentum tensor. 
One shall be reminded that (\ref{FE1}), and indeed (\ref{FE2})--(\ref{equiv}) as well
work only for the coincident gauge.
It is hence necessary to identify the covariant form for (\ref{FE1}).
It follows from  (\ref{curvature}) that we can re-express (\ref{E-1})--(\ref{FE1}) as \cite{zhao,gde}
\begin{align}\label{E-2}
E_{\mu\nu}
=&f_Q \m{G}_{\mu\nu}+\frac{1}{2} g_{\mu\nu} (f_QQ-f) + 2f_{QQ} P^\lambda{}_{\mu\nu} \m{\nabla}_\lambda Q = \kappa \Theta_{\mu\nu}.
\end{align}
It is also noticeable from the divergence of (\ref{E-2}) that
\begin{equation}
\m\nabla_\mu E^\mu{}_\nu=f_{QQ}\Phi_\nu+f_{QQQ}\Psi_\nu\,,
\end{equation}
where 
\begin{align}
\Phi_\nu=&
   \left(\m{G}^\lambda_\nu+\frac{Q}2\delta^\lambda_\nu\right)\nabla_\lambda Q
        +2\m\nabla_\mu\left( P^{\lambda\mu}_{\quad\nu}\nabla_\lambda Q\right)\,,
\end{align}
\begin{equation}
\Psi_\nu=2P^{\lambda\mu}_{\quad\nu}\m{\nabla}_\mu Q\m{\nabla}_\lambda Q.
\end{equation}
For the null divergence of the energy-momentum tensor in general, we thus must obtain
\begin{equation}\label{z1}
0=f_{QQ}\Phi_\nu+f_{QQQ}\Psi_\nu.
\end{equation}
Notice that the two vectors $\Phi_\nu$ and $\Psi_\nu$, exclusively generated from the studied spacetime geometry, are model independent and trivially vanish for a constant $Q$. Unfortunately, from the field equation (\ref{E-2}) it is clear that a constant $Q=Q_0$ reduces the present theory to GR with a cosmological constant 
\begin{align}
\Lambda=\frac12\left(Q_0-\frac{f(Q_0)}{f_Q(Q_0)} \right).
\end{align}
Therefore, in what follows we are interested only in non-constant $Q$ scenarios.

We can encounter several cases as listed next depending on the underlying spacetime geometry, and as a consequence of non-constant $Q$.
\begin{description}
\item[Case (a) $\Phi_\nu=0$, $\Psi_\nu=0$]
In this case, our lockgates are open, any $f(Q)$ model will satisfy (\ref{z1}). We have no obligation towards the viability of the models from the energy conservation ground. 
\item[Case (b) \label{caseb}$\Phi_\nu=0$, $\Psi_\nu\neq 0$]
To satisfy (\ref{z1}), only possible choice is the quadratic model $f(Q)=\alpha+\beta Q+\lambda Q^2$.
\item[Case (c) $\Phi_\nu\neq 0$, $\Psi_\nu=0$]
Clearly $f(Q)=\alpha+\beta Q$ is the only possibility. However, its dynamics is equivalent to GR with a cosmological constant $\frac{-\alpha}{2\beta}$.\label{casec}
\item[Case (d) 
 $\Phi_\nu\neq0$, $\Psi_\nu\neq0$]\label{cased}
In this particular case, we can have two scenarios as follows:
\begin{itemize}
     \item $\Phi_\nu=c\Psi_\nu,$ for some constant $c$.
This gives us a simple situation. We can integrate $0=cf_{QQ}+f_{QQQ}$ to obtain a particular model $f(Q)=\alpha+\beta Q+\lambda e^{-cQ}$.
    \item $\Phi_\nu\neq c\Psi_\nu,$ for any constant $c$, or in other sense the two vectors are linearly independent. In this case, $f(Q)=\alpha+\beta Q$, that indicates GR case with a cosmological constant.
\end{itemize}

\end{description}

To elaborate the result, in the studied spacetime let us fix a line element, that means fixing the metric along with the coordinates. Now for any flat (curvature-free) as well as symmetric (torsion-free) connection to formulate the modified $f(Q)$ gravity EoMs (equations of motion), we are dealing with a fixed choice of the two vectors $\Phi_\nu$ and $\Psi_\nu$. These two vectors can be identically zero or either of them can be non-null in the studied setting, as we show in the next section.

\subsection{Homogeneous and isotropic ansatz}
In this section, we investigate the homogeneous and isotropic standard model of cosmology. Following the cosmological principle, the Universe can be described by the Friedmann-Lema\^{i}tre-Robertson-Walker (FLRW) spacetime
\begin{align}\label{ds:RW}
ds^2=-dt^2+a^2\left(\frac{dr^2}{1-kr^2}+r^2d\theta^2+r^2\sin^2\theta d\phi^2\right)
\end{align}
where $a(t)$ is the scale factor of the Universe, $H=\frac{\dot{a}}{a}$ is the Hubble parameter and $k$ denotes the spatial curvature ($k=0,+1,-1$). $k=0$ depicts the spatially flat Universe model, i.e., the spatial part of the model is ordinary Euclidean space, whereas the $k=1$ and $k=-1$ models demonstrate the closed and open type Universe models, respectively. 

We consider all the compatible connections in this spacetime and the related analysis.
Firstly, such affine connections are required to be invariant under rotations and spatial translations due to the homogeneity and isotropy of the metric (\ref{ds:RW}). In other words, the Lie derivatives of the affine connections are required to be vanish with respect to the following Killing vector fields \cite{hohmann}  
\begin{align*}
X_x=&\sin\phi\partial_\theta+\cos\phi\cot\theta\partial_\phi; \quad
R_y=-\cos\phi\partial_\theta+\sin\phi\cot\theta\partial_\phi; \quad
R_z=-\partial_\phi\,,\\
T_x=&\chi\sin\theta\cos\phi\partial_r+\frac\chi r\cos\theta\cos\phi\partial_\theta
       -\frac\chi r\csc\theta\sin\phi\partial_\phi\,, \\
T_y=&\chi\sin\theta\sin\phi\partial_r+\frac\chi r\cos\theta\sin\phi\partial_\theta
       +\frac\chi r\csc\theta\cos\phi\partial_\phi\,, \\
T_x=&\chi\cos\theta\partial_r-\frac\chi r\sin\theta\partial_\theta\,,
\end{align*}
$\chi=\sqrt{1-kr^2}$. Furthermore, the affine connections are also demanding to meet the postulates  of symmetric teleparallelism, i.e., 
\[
R^\lambda{}_{\mu\alpha\nu}=0; \quad T^\lambda{}_{\mu\nu}=0.
\]
Having considered these postulates above, we arrive at the following class of affine connections
\cite{fQfT1, FLRW/connection}.
\begin{align} 
\Gamma^t{}_{tt}=&C_1, 
	\quad 					\Gamma^t{}_{rr}=\frac{C_2}{\chi^2}, 
	\quad 					\Gamma^t{}_{\theta\theta}=C_2r^2, 
	\quad						\Gamma^t{}_{\phi\phi}=C_2r^2\sin^2\theta,								\notag\\
\Gamma^r{}_{tr}=&C_3, 
	\quad  	\Gamma^r{}_{rr}=\frac{kr}{\chi^2}, 
	\quad		\Gamma^r{}_{\theta\theta}=-\chi^2r, 
	\quad		\Gamma^r{}_{\phi\phi}=-\chi^2r\sin^2\theta,												\notag\\
\Gamma^\theta{}_{t\theta}=&C_3, 
	\quad		\Gamma^\theta{}_{r\theta}=\frac1r,
	\quad		\Gamma^\theta{}_{\phi\phi}=-\cos\theta\sin\theta,										\notag\\
\Gamma^\phi{}_{t\phi}=&C_3, 
	\quad 	\Gamma^\phi{}_{r\phi}=\frac1r, 
	\quad 	\Gamma^\phi{}_{\theta\phi}=\cot\theta,
\end{align}
where $C_1$, $C_2$ and $C_3$ are functions of $t$, which must fulfill one of the following criteria:  
\begin{enumerate}\label{casesConn}
\item[(I)] $C_1=\gamma$, $C_2=C_3=0$ and $k=0$, where $\gamma$ is a function on $t$; or
\item[(II)] $C_1=\gamma+\dfrac{\dot\gamma}\gamma$, $C_2=0$, $C_3=\gamma$ and $k=0$,
             where $\gamma$ is a nonvanishing function on $t$; or 
\item[(III)] $C_1=-\dfrac k\gamma-\dfrac{\dot\gamma}{\gamma}$, $C_2=\gamma$, $C_3=-\dfrac k\gamma$ and $k=0,\pm1$,
             where $\gamma$ is a nonvanishing function on $t$,
\end{enumerate} 
where the $\dot{(~)}$ denotes the derivative with respect to $t$. We can then calculate the non-metricity tensor
\begin{align}
Q_{ttt}=&2C_1,
\quad Q_{tr}{}^r=Q_{t\theta}{}^{\theta}=Q_{t\phi}{}^{\phi}=-2\left(C_3-H\right),
\quad Q^r{}_{rt}=Q^{\theta}{}_{\theta t}=Q^{\phi}{}_{\phi t}=-C_3+\frac{C_2}{a^2}.
\end{align}
The tensor $P_{\lambda\mu\nu}$ can further be calculated as
\begin{align}
P_{ttt}=&\frac34\left(-C_3+\frac{C_2}{a^2}\right), 
\quad P_{tr}{}^{r}=P_{t\theta}{}^\theta=P_{t\phi}{}^\phi
				=\frac14\left(4H-3C_3-\frac{C_2}{a^2}\right) 
            , \quad 
P^r{}_{rt}=P^\theta{}_{\theta t}=P^\phi{}_{\phi t}
				=\frac14\left(C_1+C_3-H\right). 
\end{align}
So, using (\ref{Q_0}), the non-metricity scalar $Q$ is derived as
\begin{align}\label{Q}
Q(t)=3\left(
-2H^2+3C_3H+\frac{C_2}{a^2}H-(C_1+C_3)\frac{C_2}{a^2}+(C_1-C_3)C_3\right).
\end{align}
The only nontrivial vectors $\Psi_\nu$ and $\Phi_\nu$ are given respectively by 
\begin{align}\label{Psi_t}
\Psi_t=&\frac{3}{2}\left(\frac{C_2}{a^2}-C_3 \right)\dot{Q}^2\,, \\
\label{Phi_t}
\Phi_t=&\frac32\left(\frac{C_2}{a^2}H-3C_3H
+\frac{\dot C_2}{a^2}-\dot C_3
-(C_1+C_3)\frac{C_2}{a^2}+(C_1-C_3)C_3-2\frac k{a^2}
\right)\dot Q+\frac32\left(\frac{C_2}{a^2}-C_3\right)\ddot Q\,.
\end{align}

We continue to examine the three type of possible connections corresponding to the three cases 
separately as follows.

\subsubsection{Connection I: $C_1=\gamma$, $C_2=C_3=0$, $k=0$. }\label{connI}
Using (\ref{Q}), (\ref{Psi_t}) and (\ref{Phi_t}), we get 
\begin{align}
Q=&-6H^2;  \\
\Psi_t=&\Phi_t=0.\label{adconnI}
\end{align}
We conclude that there is no restriction on the function $f$ in this case, the underlying spacetime geometry carries the privilege of producing null divergence of energy-momentum tensor. 

\begin{remark}[Cartesian coordinates]\label{remark_cartesian}
Note that, for the spatially flat case $k=0$ and connection I discussed in the previous subsection \ref{connI}, we can also independently analyse the line element 
\begin{equation} \label{flrw}
ds^2 = -dt^2 + a^2(t)\delta_{ij}dx^i dx^j 
\end{equation}
in equivalent Cartesian coordinates since this is the most popular and extensively studied way, favoured by almost all researchers. In Appendix \ref{cartesian}, we present this case in greater depth and clarity than the rest of the cases, in the confidence that it will further aid future research in applying $f(Q)$ theory correctly to the field of cosmology. Since we are utilising Cartesian coordinates, rest assured that we are employing a coincident gauge, which simplifies the computations considerably. We explicitly calculate the divergence of the energy-momentum tensor without utilising the connection field equation (\ref{FE2}) and it still vanishes trivially in a model-independent manner. Indeed, this is due to (\ref{adconnI}), and this is the actual background reason why in the literature there is uninterrupted study of this specific geometry (\ref{flrw}) in cosmological sector with all kind of $f(Q)$ models.  

\end{remark}

\subsubsection{Connection II: $C_1=\gamma+\dfrac{\dot\gamma}\gamma$, $C_2=0$, $C_3=\gamma$, $k=0$. }
Using (\ref{Q}), (\ref{Psi_t}) and (\ref{Phi_t}), we obtain the non-metricity scalar $Q(t)$ and the only possible non-zero vectors $\Psi_\nu$ and $\Phi_\nu$ as 
\begin{align}
Q&=9H\gamma+3\dot\gamma-6H^2\label{qdes}\\
\Psi_t&=\frac{-3}{2}\gamma\dot{Q}^2\\
\Phi_t&=-\frac{3}{2}\gamma\left(\ddot Q+3H\dot Q\right).
\end{align}
For non-constant $Q$, $\Psi_t$ obviously cannot be zero as $\gamma$ is a non-zero function of time variable for the current connection. Hence, the only viable $f(Q)$ models are described by either case (b) or case (d) depending on $\Phi_t$.

Even if $\gamma(t)=\gamma_0$ is a constant, for vanishing $\Phi_t$, one needs $\dot{Q}=\frac C{a^3}$, $C$ a constant, and the scale factor must satisfy
\begin{align}\label{as1}
    9\gamma_0 a^2\ddot{a}-9\gamma_0a\dot{a}^2-12a\dot{a}\ddot{a}+12\dot{a}^3=C.
\end{align}
Equation (\ref{as1}) can be cast as 
\begin{align}
(12H-9\gamma_0)H^2a^3(q+1)=C,
\end{align}
where $q=-\frac{a\ddot{a}}{\dot{a}^2}$ is the deceleration parameter. Since we do not associate matter of any form while deriving (\ref{as1}), a de-Sitter solution should be viable at that epoch. Let us therefore seek a de-Sitter solution in the form $a=e^{\lambda t}$, so that $H=\lambda$. In that case, $q=-1$ requires $C=0$ which forces $\dot{Q}$ to vanish ultimately. Note that, for a constant $Q=Q=Q_0$, (\ref{qdes}) yields a quadratic equation of $H$ and its two roots are given by 
\begin{align}
\frac{9\gamma_0\pm\sqrt{81\gamma_0^2-24Q_0}}{12}.
\end{align}
Hence de-Sitter solution is admissible in a model-independent manner.   

\subsubsection{Connection III: $C_1=-\dfrac k\gamma-\dfrac{\dot\gamma}\gamma$, $C_2=\gamma$, 
$C_3=-\dfrac k\gamma, k=0,\pm 1$. }
Using (\ref{Q}), (\ref{Psi_t}) and (\ref{Phi_t}), we calculate the non-metricity scalar $Q(t)$ to be
\begin{equation}
Q=-3\left\{\left(2H+\frac{3k}\gamma-\frac\gamma{a^2}\right)H-\frac{2k}{a^2}-\frac{k\dot\gamma}{\gamma^2}-\frac{\dot\gamma}{a^2}\right\}.
\end{equation}
The only possible non-zero vector $\Psi_\nu$ in this scenario is
\begin{align}
    \Psi_t=\frac{3}{2}\left[ \frac{k}{\gamma}+\frac{\gamma}{a^2} \right]\dot{Q}^2
\end{align}
which for non-constant $Q$, is definitely non-zero for $k=1$ case as it gives us imaginary function $\gamma(t)$. If $k=-1$, $\Psi_t$ vanishes if $\gamma(t)^2=a^2(t)$.
Whereas, the only possible non-zero $\Phi_\nu$ is
\begin{align}
\Phi_t=\frac{3}{2}\left(\frac\gamma{a^2}H+3\frac k\gamma H+2\frac{\dot \gamma}{a^2}\right)\dot{Q}
    +\frac{3}{2}\left(\frac{k}{\gamma}+\frac{\gamma}{a^2}\right)\ddot{Q}.
\end{align}
The aforementioned rationale leaves us with no choice but to evaluate the case when $k=-1$, and $\gamma=\pm a$. Remarkably, under this condition, $\Phi_t$ vanishes even for non-constant $Q$. 

We conclude that in a spatially open ($k=-1$) FLRW Universe, under a specific choice of $\gamma(t)=\pm a(t)$ for connection III, $\Phi_t=0=\Psi_t$ which implies that the energy conservation criterion is satisfied for any $f(Q)$ model. So we have no restriction imposed on the viability of models from this ground. However, this is not true in general for any connection III (as $\gamma(t)$ can be arbitrarily chosen to yield more such connections). On the other hand, for spatially flat ($k=0$) and spatially closed ($k=1$) FLRW Universe, $\Phi_t=0=\Psi_t$ demands a constant $Q$. For a non-constant $Q$, the viable models fall into one of three categories: linear, quadratic, or exponential in $f(Q)$, as discussed in the Case (b)-Case (d).    
\begin{remark}
    To further convince the readers about the importance of our analysis, let us recall that in FLRW spacetime we so frequently use the continuity equation (derived from a vanishing $\mathring{\nabla}^\mu T_{\mu\nu}$, for a perfect fluid type energy-momentum tensor $T_{\mu\nu}=(p+\rho)u_\mu u_\nu+pg_{\mu\nu}$) 
    $$
    \dot{\rho}=-3H(p+\rho)=-3H(1+\omega)\rho\,,
    $$
which on integration yields $\rho=\rho_0 a^{-3(1+\omega)}$. Although for connection I in spatially flat case and for connection III in open type Universe ($k=-1$) we can continue with such major information about the energy density; in other cases we can not use the continuity equation trivially for just any $f(Q)$ models.

\end{remark}

\subsection{Static and spherically symmetric ansatz}
As another interesting example in this regard, we consider the spherically symmetric solution of $f(Q)$-gravity first discussed in \cite{lin}. We begin with the metric
\begin{equation}\label{ss}
ds^2=-e^{\xi(r)} dt^2+e^{\zeta(r)} dr^2+r^2d\Omega^2.
\end{equation} 
Then, by cancelling gravity to derive the required symmetric teleparallel affine connection coefficients as prescribed in \cite{lin}, we proceed to compute the only non-zero component of $\Psi_\nu$ to be
\begin{equation}
\Psi_r=\frac{e^{-\zeta}-1}{r}\left(Q'\right)^2\,,
\end{equation}
which is non-zero, in general, unless we force additional constraints (like $\zeta=0$ or $Q'=0$) on the metric function to make it null. 
 Here ${(~)}'$ denotes the derivative over $r$.
However, such a condition, once enforced, pushes the non-metricity scalar 
\begin{align}
Q(r)=\frac{\left(e^{-\zeta} - 1\right)\left(\xi' +\zeta'\right)} {r}
\,,
\end{align}
to be a constant, or even worse make it zero (for vanishing $\zeta$) and the essence of the modified $f(Q)$ theory is lost. 
The only non-zero vector $\Phi_\nu$ in this case is given by
\begin{equation}\Phi_r= \frac{r\left(e^{-\zeta}-1\right) \xi'
 - r\left(e^{-\zeta}+1\right) \zeta' 
 + 4 \, \left(e^{-\zeta}-1\right)}{2 \, r^{2}}Q'
+\frac{e^{-\zeta}-1}{r}Q''.
\end{equation}
The above restriction (either $\zeta=0$ or $Q'=0$) arises for vanishing $\Psi_r$, also forces $\Phi_r=0$. 

The authors of \cite{lin} ultimately also arrived to the conclusion ($Q=0$) for static spherically symmetric vacuum solution in $f(Q)$ theory, from some separate argument based on the equations of motion. Additionally it appeared that 
$\xi'+\zeta'=0$ and thus an exactly Schwarzschild-de Sitter like general relativistic solution with cosmological constant (only the cosmological constant depending on the $f(Q)$ model) was recovered.     


\subsubsection{Wormhole solution}
As a special case of static spherically symmetric spacetime, let us mention the wormhole (WH) metric 
\begin{align}ds^2 = -e^{\xi(r)} 
dt^2 +  \frac{1}{1-\frac{b(r)}r}
dr^2 + r^2 d\Omega^2. 
\end{align}
$b(r)$ is known as the wormhole shape function as it specifies the geometrical shape of the wormhole. This metric can be obtained by replacing $e^{-\zeta}$ by $1-\frac br$ in (\ref{ss}). The non-metricity scalar is given by
\begin{align}
    Q=-\frac b{r^2}\left[\frac{rb'-b}{r(r-b)}+\xi'\right].
\end{align}
We can compute the only possibly non-zero vector $\Psi_\nu$ in the similar manner 
\begin{align}
    \Psi_r= -\frac{b}{2r^2}\left(Q'\right)^2.
\end{align}
For a vanishing $\Psi_r$, one needs either $b=0$ or $Q'=0$. However, $b=0$ is not physically admissible in a wormhole geometry. Therefore, $Q$ must be a constant for the $f(Q)$ models to be unrestricted upon the energy conservation criteria in wormhole geometry.

The other important nonzero vector component $\Phi_\nu$ in this case is
\begin{align}
    \Phi_r= -\frac{rb(r-b)\xi'+r(2r-b)b'+b(2r-3b)}{2r^3(r-b)}Q'-\frac{b}{r^2}Q''\,,
\end{align}
and it vanishes as well for a constant $Q$.

\section{\textbf{Concluding remarks}}\label{sec5}
In the present study we have made a novel attempt to demystify the three interconnecting aspects of $f(Q)$ theory - the classical energy conservation criterion, the two independent field equations of $f(Q)$ theory generated by varying the metric and connection, and the extended Bianchi identity. 
We have shown that the connection field equation and the energy conservation criterion are actually equivalent condition in this theory. More importantly, neither of these two conditions are true in generic $f(Q)$ theory, unless $Q$ is constant. 
This vital feature has been predominantly overlooked in the literature, most probably since the condition holds in spatially flat FLRW metric in Cartesian coordinates (coincident gauge choice), as proved explicitly in the present discussion.
Thus for the first time, the hazy pictures of why the connection field equation can be wiped out clean from all the discussions in this particular cosmological setting, or why we can employ the continuity equation $\dot{\rho}+3H(p+\rho)=0$ in this case, are explained with a great precision.
We have further shown that the said criterion is not always satisfied in FLRW background (spatially flat and open or closed type) with some other affine connection and frame choice.
The matter of unfortunate fact is that the spatially flat FLRW metric in Cartesian coordinates (coincident gauge) cannot let us distinguish between the metric and symmetric teleparallel theories. 
Therefore, we cannot stick to this special coordinate and gauge choice in FLRW if we honestly like to invest in this newly proposed $f(Q)$ theories of gravity in demonstrating the unexplained observations of our Universe. 
And as soon we put our step in the other possible connections and frames in FLRW geometry, the energy conservation criterion/connection's field equation are not satisfied anymore for general $f(Q)$ models. 

On the other hand, the most important model of the astrophysical objects, the static spherically symmetric solution in $f(Q)$ gravity, so far rather remains unclear.
Although BH solution was attempted \cite{lin}, it ended with constant non-metricity scalar $Q=0$. At the background level such solutions are the same as in general relativity with a cosmological constant. 
This is a temporary success where the field equation is heavily reduced by considering a constant $Q$. Very recently, in a wonderful work \cite{bh} some beyond-GR BH solutions have been described including the precise conditions under which these can be obtained. The respective classes of connections are different from that used in \cite{lin} and whether such connections can pass the energy conservation criterion, is a future research question.   
Model-specific WH solutions were discussed in $f(Q)$ theory \cite{wh1,wh2}, however further research on this topic is still needed. \cite{wh1} mentioned the non-existence of WH for quadratic $f(Q)$ model, but no further attempt on investigating the actual reason of this discrepancy was done.   

And here comes our crucial study, showing that the root cause lies in the fact that the energy conservation criterion/connection field equation/extended Bianchi identity, whatever terminology we impose, is not always satisfied in $f(Q)$ theory, for non-constant $Q$. It is rather impossible to obtain a beyond GR $f(Q)$ model in this modified theory which is consistent with the energy conservation or equivalently the connection's field equation in every spacetime geometry.

\appendix

\section{Appendix: Calculations regarding equation (\ref{equiv})}\label{appEquiv}
We shall verify (\ref{equiv}) in this appendix.
It suffices to verify the following relation in the coincident gauge choice:
\begin{align}
\frac2{\sqrt{-g}}\partial_\lambda\partial_\alpha\left(\sqrt{-g}f_QP^{\alpha\lambda}{}_{\nu}\right)
=\m\nabla_\mu E^\mu{}_\nu\,.
\end{align}
Firstly, recalling that 
\begin{align*}
\partial_\mu \sqrt{-g}=\frac12\sqrt{-g}Q_\mu=-\sqrt{-g}\tilde L_\mu\,,
\end{align*}
were $\tilde L_\mu=L^\lambda{}_{\mu\lambda}.$
It follows that 
\begin{align*}
\partial_\alpha\left(\sqrt{-g}f_QP^{\alpha\lambda}{}_{\nu}\right)
=&\sqrt{-g}\left(f_{QQ}P^{\alpha\lambda}{}_\nu\nabla_\alpha Q
			+f_Q(\nabla_\alpha-\tilde L_\alpha)P^{\alpha\lambda}{}_\nu\right)\,,
\end{align*}
and so
\begin{align*}
\frac1{\sqrt{-g}}\partial_\lambda\partial_\alpha\left(\sqrt{-g}f_QP^{\alpha\lambda}{}_{\nu}\right)
=&f_{QQQ}A_{3,\nu}+f_{QQ}A_{2,\nu}+f_QA_{1,\nu}\,,
\end{align*}
where 
\begin{align*}
A_{1,\nu}=&(\nabla_\lambda-\tilde L_\lambda) (\nabla_\alpha-\tilde L_\alpha) P^{\alpha\lambda}{}_\nu\,, \\
A_{2,\nu}=&(\nabla_\lambda-\tilde L_\lambda)(P^{\alpha\lambda}{}_\nu\nabla_\alpha Q)+
     (\nabla_\alpha-\tilde L_\alpha)P^{\alpha\lambda}{}_\nu\cdot\nabla_\lambda Q\,, \\
A_{3,\nu}=&P^{\alpha\lambda}{}_\nu\nabla_\alpha Q\nabla_\lambda Q	\,.	
\end{align*}
Clearly $2A_{3\nu}=\Psi_\nu$.
Next we verify that $2A_{2,\nu}=\Phi_\nu$ and $A_{1,\nu}=0$.
For this purpose, we first prepare the following relations for later use.
\begin{align}
2\left(\nabla_\alpha-\tilde L_\alpha\right)P^{\alpha\lambda}{}_\nu
=&2\left(\m\nabla_\alpha P^{\alpha\lambda}{}_\nu+L^\lambda{}_{\beta\alpha}P^{\alpha\beta}{}_\nu
												                 -L^\beta{}_{\nu\alpha}P^{\alpha\lambda}{}_\beta\right) \notag\\
=&\m G^\lambda{}_\nu+\frac Q2\delta^\lambda{}_\nu+2L^\beta{}_{\nu\alpha}P^{\lambda\alpha}{}_\beta				\label{150a}\\
=&\m G^\lambda{}_\nu+\frac Q2\delta^\lambda{}_\nu-Q_{\nu\alpha\beta}P^{\lambda\alpha\beta}\,. 		\label{150b}
\end{align}
Following the same calculation as in deriving the variation of $Q$ \cite{fQT}, gives
\begin{align}
\nabla_\nu Q-2P^{\sigma\alpha\rho}\nabla_\sigma Q_{\nu\alpha\rho}
=&\left(-P^{\alpha\sigma\rho}Q^\beta{}_{\sigma\rho}
			-2P^{\sigma\rho\alpha}Q_{\sigma\rho}{}^\beta\right)Q_{\nu\alpha\beta}	\notag\\
=&\left(P^{\alpha\sigma\rho}L_{\rho\sigma\beta}
					+2P^{\sigma\rho\alpha}(L_{\rho\sigma\beta}+L_{\beta\rho\sigma})\right)Q_{\nu\alpha}{}^\beta\,,
\label{160}
\end{align}

\begin{align*}
2A_{2,\nu}=&2\m\nabla_\lambda(P^{\alpha\lambda}{}_\nu\nabla_\alpha Q)+
     2\left(-L^\beta{}_{\nu\alpha}P^{\lambda\alpha}{}_\beta+
					(\nabla_\alpha-\tilde L_\alpha)P^{\alpha\lambda}{}_\nu\right)\nabla_\lambda Q 		\\
  =&\Phi_\nu\,,
\end{align*}
to which we have applied (\ref{150a}).
Finally we compute 
\begin{align*}
2A_{1,\nu}
=&\m\nabla_\lambda\left(\m G^\lambda{}_\nu+\frac Q2\delta^\lambda{}_\nu\right)
		-L^\beta{}_{\nu\lambda}\left(\m G^\lambda{}_\beta+\frac Q2\delta^\lambda{}_\beta\right)
		-\nabla_\lambda Q_{\nu\alpha\beta}\cdot P^{\lambda\alpha\beta}
		-Q_{\nu\alpha\beta}(\nabla_\lambda-\tilde L_\lambda)P^{\lambda\alpha\beta} \\
=&\frac12\nabla_\nu Q-\nabla_\lambda Q_{\nu\alpha\beta}\cdot P^{\lambda\alpha\beta}
		-L^\beta{}_{\nu\lambda}\left(\m G^\lambda{}_\beta+\frac Q2\delta^\lambda{}_\beta\right)
		-Q_{\nu\lambda}{}^\beta(\nabla_\sigma-\tilde L_\sigma)P^{\sigma\lambda}{}_\beta
		+Q_{\nu\lambda\alpha}P^{\sigma\lambda}{}_\beta Q_\sigma{}^{\beta\alpha}\\
=&0\,,
\end{align*}
where we have applied the fact that $[\nabla_\lambda,\nabla_\nu]g_{\alpha\beta}=0$ (due to the flatless of the affine connection) and (\ref{160}) to the first two terms; and (\ref{150b}) to the remaining terms in the second equality.

\section{Appendix: Calculations regarding Remark \ref{remark_cartesian} }\label{cartesian}
With the notations $u_\mu=(\partial_t)_\mu, H(t)=\frac{\dot a}{a}$; we have the following expressions \cite{Avik/prd}
\begin{align}
\m\nabla_\mu u_{\nu}=&H(g_{\mu\nu}+u_{\mu}u_{\nu})\,,                           \label{eqn:a00}\\
\m G_{\mu\nu}=&-(2\dot H+3H^2)g_{\mu\nu}-2\dot Hu_\mu u_\nu\,,                  \label{eqn:einstein}\\
Q=&-6H^2\,,                                                                     \label{eqn:a10}\\
\nabla_\lambda Q=&- \dot Qu_\lambda=12H\dot H u_\lambda\,,                      \label{eqn:a20}\\
\nabla_\lambda QP^\lambda{}_{\mu\nu}=&12H^2\dot H (g_{\mu\nu}+u_{\mu}u_{\nu}).  \label{eqn:a30}
\end{align}
Using these data, we can express (\ref{E-2}) as 
\begin{align} \label{FE-2}
f_Q \m{G}^\mu_\nu+\left(-3f_QH^2-\frac{1}{2}f \right)\delta^\mu_\nu
 + 24f_{QQ}H^2\dot H(\delta^\mu_\nu+u^\mu u_\nu )=\kappa \Theta^\mu_\nu. 
 \end{align}
Now the divergence of (\ref{FE-2}) gives
\begin{align} \label{eqn:a40}
\kappa\m \nabla_\mu \Theta^\mu_\nu		=& X_1+X_2+X_3+X_4\,,
\end{align}
where the vectors
\begin{align*} 
X_1=& \m{G}^\mu_\nu\nabla_\mu f_Q\,,                     \\
X_2=&\nabla_\nu\left(-3f_QH^2-\frac{1}{2}f \right)\,,    \\
X_3=&24f_{QQ}H^2\dot H \m\nabla_\mu(u^\mu u_\nu )\,,     \\
X_4=&24(\delta^\mu_\nu+u^\mu u_\nu )\nabla_\mu(f_{QQ}H^2\dot H).
\end{align*}
We proceed to compute each of these $X_1,\,X_2,\,X_3,\,X_4$.

Using  (\ref{eqn:a20}), we obtain
\[
\nabla_\mu f_{QQ}=f_{QQQ}\nabla_\mu Q=12f_{QQQ}H\dot H u_\mu.
\]
Hence 
\[
(\delta^\mu_\nu+u^\mu u_\nu )\nabla_\mu f_{QQ}
=12f_{QQQ}H\dot H(\delta^\mu_\nu+u^\mu u_\nu )u_\mu=0.
\]
Similarly
\[
(\delta^\mu_\nu+u^\mu u_\nu )\nabla_\mu H= (\delta^\mu_\nu+u^\mu u_\nu )\nabla_\mu \dot H=0.
\]
Hence 
\begin{align}\label{eqn:X4b} X_4=0.
\end{align}
Next, applying (\ref{eqn:einstein}) and (\ref{eqn:a20}), we get
\begin{align}\label{eqn:X1b}
X_1=&
12f_{QQ}H\dot H\big\{-(2\dot H+3H^2)\delta^\mu_\nu-2\dot Hu^\mu u_\nu\big\}u_\mu \nonumber\\
=&-36f_{QQ}H^3\dot Hu_\nu.
\end{align}
Similarly
\begin{align}\label{eqn:X2b}
X_2=&-3f_{QQ}H^2\nabla_\nu Q+6f_QH\dot Hu_\nu-\frac{1}{2}f_Q\nabla_\nu Q \nonumber\\
=&-36f_{QQ}H^3\dot H u_\nu. 
\end{align}
Using (\ref{eqn:a00}) we obtain
\begin{align} \label{eqn:X3b}
X_3
=&24f_{QQ}H^2\dot H(u_\nu \m\nabla_\mu u^\mu+u^\mu\m\nabla_\mu u_\nu )\nonumber\\
=&72f_{QQ}H^3\dot H.
\end{align}
It follows from (\ref{eqn:X4b})--(\ref{eqn:X3b}) that 
\[
\m\nabla_\mu\Theta^\mu_\nu=X_1+X_2+X_3+X_4=0.
\]


\begin{thebibliography}{99}
\bibitem{einst} 
A. Unzicker and T. Case, Translation of Einstein's Attempt of a Unified Field 
Theory with Teleparallelism, arXiv:physics/0503046 (2005).

\bibitem{nester} J.M. Nester, H-J Yo, Symmetric teleparallel general relativity, Chin.J.Phys. \textbf{37}, 113 (1999). 

\bibitem{coincident} 
J. B. Jimenez, L. Heisenberg and T. Koivisto, Coincident General Relativity, 
Phys. Rev. D, \textbf{98}, 044048 (2018).

\bibitem{f(T)} R. Ferraro and F. Fiorini, Modified teleparallel grav-
ity: Inflation without inflaton, Phys. Rev. D \textbf{75}, 084031
(2007).

\bibitem{accfT1} 
G. R. Bengochea and R. Ferraro, 
Dark torsion as the cosmic speed-up, 
Phys. Rev. D, \textbf{79}, 124019 (2009).

\bibitem{accfT2} 
E. V. Linder, Einstein's Other Gravity and the Acceleration of the Universe, 
Phys. Rev. D, \textbf{81}, 127301 (2010).

\bibitem{accfT3} 
K. Bamba, C. Q. Geng, C. C. Lee and L. W. Luo, Equation of state for dark energy in $f(T)$ gravity, 
JCAP, \textbf{2011}(01), 021 (2011).


\bibitem{accfT4} 
K. Bamba, C. Q. Geng and C. C. Lee, 
Comment on ``Einstein's Other Gravity and the Acceleration of the Universe", 
arXiv:1008.4036 [astro-ph.CO] (2010).

\bibitem{dy2} 
S. A. Narawade, L. Pati, B. Mishra and S. K. Tripathy, Dynamical system analysis for accelerating models in non-metricity $f(Q)$ gravity, 
Phys. Dark Univ., \textbf{36}, 101020 (2022).

\bibitem{accfQ1} 
R. Solanki, A. De, S. Mandal and P. K.Sahoo, Accelerating expansion of the universe in modified symmetric teleparallel gravity, 
Phys. Dark Univ., \textbf{36}, 101053 (2022).

\bibitem{accfQ2} 
R. Solanki, A. De and P. K. Sahoo, Complete dark energy scenario in $f(Q)$ gravity, 
Phys. Dark Univ.,  \textbf{36}, 100996 (2022).

\bibitem{accfQ3} 
L. Atayde and N. Frusciante, Can $f(Q)$ gravity challenge $\Lambda$CDM?, 
Phys. Rev. D,  \textbf{104}, 064052 (2021).

\bibitem{fQfT} 
J. B. Jimenez, L. Heisenberg and T.S. Koivisto, The Geometrical Trinity of Gravity, 
Universe,  \textbf{5}, 173 (2019).

\bibitem{fQfT1}
F. D'Ambrosio, L. Heisenberg and S. Kuhn, Revisiting Cosmologies in Teleparallelism, 
Class. Quantum Grav., \textbf{39}, 025013 (2022).

\bibitem{fQfT2} 
J. Lu, Y. Guo and G. Chee, From GR to STG - Inheritance and development of Einstein's heritages, arXiv:2108.06865 (2021).

\bibitem{fQfT3} 
S. Capozziello, V. De Falco and C. Ferrara, Comparing Equivalent Gravities: common features and differences, 
arXiv:2208.03011 [gr-qc] (2022).

\bibitem{conserve}
H. Velten and T.R.P. Carames, 
To Conserve, or Not to Conserve:\, A Review of Nonconservative Theories of Gravity,
 Universe, \textbf{7} (2021) 38.

\bibitem{conserve1} 
T. Koivisto, Covariant conservation of energy momentum in modified gravities, 
Class. Quant. Grav., \textbf{23} (2006) 4289--4296.


\bibitem{conserve2}
T. Harko, T. S. Koivisto, F. S. N. Lobo, G. J. Olmo and D. R. Garcia, 
Coupling matter in modified $Q$-gravity, 
Phys. Rev. D, \textbf{98}, 084043 (2018).

\bibitem{conserve3} 
J. B. Jimenez, L. Heisenberg and T. S. Koivisto, 
Teleparallel Palatini theories, 
JCAP, \textbf{2018}(08), 039 (2018)


\bibitem{cosmoconserve} 
J. B. Jimenez, L. Heisenberg, T. S. Koivisto and S. Pekar, 
Cosmology in $f(Q)$ geometry, 
Phys. Rev. D, \textbf{101}, 103507 (2020).



\bibitem{bianchift}
A. Golovnev and M.J. Guzm\'an, 
Bianchi identities in f(T)gravity: Paving the way to confrontation with astrophysics, 
Phys. Lett. B,   \textbf{810},  135806 (2020).

\bibitem{bianchift2}
A. Golovnev and T. Koivisto, 
Cosmological perturbations in modified teleparallel gravity models,
JCAP, \textbf{2018}(11), 012 (2018)


\bibitem{fT/issue} 
A. Golovnev and M-J Guzm\'{a}n, 
Foundational issues in $f(T)$ gravity theory,  
Int. J. Mod. Phys. D, \textbf{18}, 2140007 (2021).


\bibitem{fT/issue2} 
B. Li, T. P. Sotiriou and J. D. Barrow, 
$f(T)$ gravity and local Lorentz invariance, 
Phys. Rev. D, \textbf{83}, 064035 (2011).

\bibitem{fT/lli} 
A. Golovnev, Issues of Lorentz-invariance in $f(T)$ gravity and calculations for spherically symmetric solutions, 
Class. Quantum Grav., \textbf{38}, 197001 (2021).

\bibitem{pereira}
R. Aldrovandi and J. G. Pereira,
\textit{Teleparallel Gravity: An Introduction},
Fundamental Theories of Physics,  Dordrecht: Springer 
\textbf{173} (2013).

\bibitem{krssak}
M. Kr\v{s}\v{s}\'{a}k and E. N. Saridakis,
{The covariant formulation of $f(T)$ gravity},
Class. Quant. Grav., \textbf{33},  115009 (2016).

\bibitem{krssak2}   
M. Kr\v{s}\v{s}\'{a}k, R. J. van den Hoogen, J. G. Pereira, C. G. B\"{o}hmer and 
A. A. Coley,
{Teleparallel Theories of Gravity: Illuminating a Fully Invariant Approach},
Class. Quant. Grav., \textbf{36},  183001 (2019).

\bibitem{bad-tetrad}
N. Tamanini and C. G. B\"ohmer,
Good and bad tetrads in $f(T)$  gravity,
Phys. Rev. D, \textbf{86},   044009 (2012).


\bibitem{weinberg}
S. Weinberg,
\textit{Gravitation and Cosmology: Principles and Applications of the General Theory of Relativity},
John Wiley \& Sons, New York (1972).

\bibitem{blagojevi}
M. Blagojevi\'c and I. A. Nikoli\'c
\textit{Hamiltonian structure of the teleparallel formulation of general relativity},
Phys. Rev. D, \textbf{62},  024021 (2000).

\bibitem{lcdm} 
F. K. Anagnostopoulos, S. Basilakos, and E. N. Saridakis, 
First evidence that non-metricity $f(Q)$ gravity can challenge $\Lambda$CDM, 
Phys. Lett. B,  \textbf{822},  136634 (2021).

\bibitem{deepjc} 
A. De, S. Mandal, J. T. Beh, T. H. Loo and P.K. Sahoo, 
Isotropization of locally rotationally symmetric Bianchi-I universe in $f(Q)$-gravity, 
Eur. Phys. J. C, \textbf{82}, 72 (2022).

\bibitem{zhao}
D. Zhao, Covariant formulation of $f(Q)$ theory, 
Eur. Phys. J. C, \textbf{82}, 303 (2022).

\bibitem{gde} 
B. J. Theng, T. H. Loo and A. De, Geodesic Deviation Equation In $f(Q)$-Gravity, 
Chinese J. of Phys., \textbf{77}, 1551 (2022).

\bibitem{lin} 
R. H. Lin and X. H. Zhai, Spherically symmetric configuration in $f(Q)$ gravity, 
Phys. Rev. D, \textbf{103}, 124001 (2021).

\bibitem{cosmography} 
S. Mandal, D. Wang and P. K. Sahoo, Cosmography in $f(Q)$ gravity, 
Phys. Rev. D, \textbf{102}, 124029 (2020).

\bibitem{signa}
N. Frusciante, Signatures of $f(Q)$-gravity in cosmology, 
Phys. Rev. D, \textbf{103}, 0444021 (2021).

\bibitem{redshift}
B. J. Barros, T. Barreiro1, T. Koivisto and N. J. Nunes, Testing $F(Q)$ gravity with redshift space distortions, 
Phys. Dark Univ., \textbf{30}, 100616 (2020).

\bibitem{perturb} 
W. Khyllep, A. Paliathanasis and J. Dutta, Cosmological solutions and growth index of matter perturbations in $f(Q)$ gravity, 
Phys. Rev. D, \textbf{103}, 103521 (2021).


\bibitem{dynamical1} 
J. Lu, X. Zhao and G. Chee, Cosmology in symmetric teleparallel gravity and its dynamical system, 
Eur. Phys. J. C, \textbf{79}, 530 (2019).

\bibitem{dynamical2} W. Khyllep, J. Dutta, E. N. Saridakis, K. Yesmakhanova, Cosmology in $f(Q)$ gravity: A unified dynamical system analysis at background and perturbation levels,  	arXiv:2207.02610 [gr-qc].

\bibitem{latetime}A. Lymperis, Late-time cosmology with phantom dark-energy in $f(Q)$ gravity, JCAP \textbf{2022}(11), 018 (2022). 

\bibitem{quantum}N. Dimakis, A. Paliathanasis and T. Christodoulakis, Quantum cosmology in $f(Q)$ theory, \textbf{38}, 225003 (2021). 

\bibitem{bouncing} F. Bajardi, D. Vernieri, S. Capozziello, Bouncing cosmology in $f(Q)$ symmetric teleparallel gravity,  Eur. Phys. J. Plus, \textbf{135}, 912 (2020). 

\bibitem{bigbang}F. K. Anagnostopoulos, V. Gakis, E. N. Saridakis, S. Basilakos, New models and Big Bang Nucleosynthesis constraints in $f(Q)$ gravity, arXiv:2205.11445 [gr-qc].

\bibitem{Avik/prd} 
A. De and T. H. Loo, Comment on ``Energy conditions in $f(Q)$ gravity", 
Phys. Rev. D, \textbf{106}, 048501 (2022).
\bibitem{FLRW/connection1}N. Dimakis, M. Roumeliotis, A. Paliathanasis, P.S. Apostolopoulos, T. Christodoulakis, Self-similar Cosmological Solutions in Symmetric Teleparallel theory: Friedmann-Lema\^itre-Robertson-Walker spacetimes, arXiv:2210.10295 [gr-qc].

\bibitem{tolman}
R. C. Tolman,  On the use of the energy-momentum principle in general relativity,
Phys. Rev. 35, 875 (1930).



\bibitem{hohmann}   
M. Hohmann,
{Variational principles in teleparallel gravity theories},
universe, \textbf{7}, 114 (2021).



\bibitem{FLRW/connection} 
N. Dimakis, A. Paliathanasis, M. Roumeliotis, and T. Christodoulakis,
{FLRW solutions in $f(Q)$ theory: The effect of using different connections},
Phys. Rev. D, \textbf{106}, 043509 (2022).

\bibitem{bh}Fabio D'Ambrosio, Shaun D.B. Fell, Lavinia Heisenberg and Simon Kuhn, 
Black holes in $f(Q)$ Gravity, Phys. Rev. D, \textbf{105}, 024042 (2022).

\bibitem{wh1}
A. Banerjee, A. Pradhan, T. Tangphati and F. Rahaman, Wormhole geometries in $f(Q)$ gravity and the energy conditions, 
Eur. Phys. J. C,  \textbf{81}, 1031 (2021).

\bibitem{wh2}
Z. Hassan, S. Ghosh, P. K. Sahoo and K. Bamba, Casimir Wormholes in Modified Symmetric Teleparallel Gravity, arXiv:2207.09945 [gr-qc].

\bibitem{fQT}
Y. Xu, G. Li, T. Harko and S. D. Liang,
{$f(Q,T)$ gravity},
Eur. Phys. J. C, \textbf{79}, 708 (2019).


\end{thebibliography}
\end{document}